# Proportionate Adaptive Filtering for Block-sparse System Identification

Jianming Liu, *Student Member, IEEE* and Steven L. Grant[1], *Member, IEEE*

*Abstract*—In this paper, a new family of proportionate normalized least mean square (PNLMS) adaptive algorithms that improve the performance of identifying block-sparse systems is proposed. The main proposed algorithm, called block-sparse PNLMS (BS-PNLMS), is based on the optimization of a mixed $l_{2,1}$ norm of the adaptive filter's coefficients. It is demonstrated that both the NLMS and the traditional PNLMS are special cases of BS-PNLMS. Meanwhile, a block-sparse improved PNLMS (BS-IPNLMS) is also derived for both sparse and dispersive impulse responses. Simulation results demonstrate that the proposed BS-PNLMS and BS-IPNLMS algorithms outperformed the NLMS, PNLMS and IPNLMS algorithms with only a modest increase in computational complexity.

## I. INTRODUCTION

SPARSE system identification has attracted much attention in the field of adaptive algorithms. The family of proportionate algorithms exploits this sparseness of a given system to improve the convergence performance of normalized least mean square (NLMS) [1]-[13] and is widely used in network echo cancellation (NEC), etc.

The idea behind PNLMS is to update each coefficient of the filter independently by adjusting the adaptation step size in proportionate to the estimated filter's coefficient [2]. The proportionate NLMS (PNLMS), as compared to the NLMS, has very fast initial convergence and tracking when the echo path is sparse. However, large coefficients converge quickly (fast initial convergence) at the cost of dramatically slowing the convergence of the small coefficients (after the initial period) [3]-[4]. As the large taps adapt, the remaining small coefficients adapt at a rate slower than NLMS.

The mu-law PNLMS (MPNLMS) algorithm proposed in [3]-[4] addresses the issue of assigning too large of an update gain to the large coefficients. The total number of iterations for overall convergence is minimized when all of the coefficients reach the $\varepsilon$-vicinity of their true values simultaneously (where $\varepsilon$ is some small positive number). The $\varepsilon$-law PNLMS (EPNLMS) algorithm is the second implementation of the same philosophy used to generate the MPNLMS algorithm [5]. The EPNLMS algorithm gives the minimum gain possible to all of the coefficients with a magnitude less than $\varepsilon$. This is based on the assumption that the impulse response is sparse and contains many small magnitude coefficients. However, the MPNLMS algorithm's performance is more robust than the EPNLMS algorithm regarding the choice of algorithm parameters, as well as input signal and unknown system characteristics [1]. Furthermore, the $l_0$ norm family algorithms have recently become popular for sparse system identification. A new PNLMS algorithm based on the $l_0$ norm was proposed to represent a better measure of sparseness than the $l_1$ norm in a PNLMS-type algorithm [6]. Benesty demonstrated that PNLMS could be deduced from a basis pursuit perspective [7]. A more general framework was further proposed to derive proportionate adaptive algorithms for sparse system identification, which employed convex optimization [8].

In many simulations, however, it seems that we fully benefit from PNLMS only when the impulse response is close to a delta function [9]. Indeed, PNLMS converges much slower than NLMS when the impulse response is dispersive. The PNLMS++ algorithm, which achieves improved convergence by alternating between NLMS and PNLMS each sample period, was proposed in an attempt to address this problem [9]. The improved PNLMS (IPNLMS) was proposed to exploit the "proportionate" idea by introducing a controlled mixture of proportionate (PNLMS) and non-proportionate (NLMS) adaptations [10]. The IPNLMS algorithm performs better than both the NLMS and the PNLMS algorithms regardless of the impulse response's nature. The improved IPNLMS (IIPNLMS) algorithm was proposed to identify active and inactive regions of the echo path impulse response [11]. Active regions receive updates that are more in-line with NLMS, while inactive regions received gains based upon PNLMS. Meanwhile, a

This work was performed under the Wilkens Missouri Endowment. Both J. Liu, and S. L. Grant are with the Electrical and Computer Engineering Department, Missouri University of Science and Technology, Rolla, MO, 65409 USA (e-mail: sgrant@mst.edu).
[1] Formerly Steven L Gay.

partitioned block improved proportionate NLMS (PB-IPNLMS) algorithm exploits the properties of an acoustic enclosure where the early path (i.e., direct path and early reflections) of the acoustic echo path is sparse and the late reverberant part of the acoustic path is dispersive [12]. The PB-IPNLMS consists of two time-domain partitioned blocks, such that different adaptive algorithms can be used for each part.

The standard PNLMS algorithm performance depends on some predefined parameters controlling proportionality through a minimum gain that is common for all of the coefficients. The individual activation factor PNLMS (IAF-PNLMS) algorithm was proposed to use a separate time varying minimum gain for each coefficient, which is computed in terms of both the past and the current values of the corresponding coefficient magnitude, and does not rely on either the proportionality or the initialization parameters [13].

The family of zero-point attracting projection (ZAP) algorithms was recently proposed to solve the sparse system identification problem [14]-[17]. When the solution is sparse, the gradient descent recursion will accelerate the convergence of the sparse system's near-zero coefficients. A block-sparsity-induced adaptive filter, called block-sparse LMS (BS-LMS), was recently proposed to improve the identification of block-sparse systems [18]. The basis of BS-LMS is to insert a penalty of block-sparsity (a mixed $l_{2,0}$ norm of adaptive tap-weights with equal group partition sizes) into the cost function of the traditional LMS algorithm.

A family of proportionate algorithms is proposed here for block-sparse system identification, which can achieve faster convergence in the block-sparse application. Both the classical NLMS and the PNLMS algorithms are special cases of this proposed scheme. The computational complexities of the proposed BS-PNLMS and BS-IPNLMS algorithms are also compared to NLMS, PNLMS, and IPNLMS algorithms.

## II. REVIEW OF PNLMS

The input signal $\mathbf{x}(n)$ is filtered through the unknown coefficients, $\mathbf{h}(n)$, so that the observed output signal $d(n)$ can be obtained as

$$d(n) = \mathbf{x}^T(n)\mathbf{h}(n) + v(n), \quad (1)$$

where

$$\mathbf{x}(n) = [x(n), x(n-1), \cdots, x(n-L+1)]^T,$$
$$\mathbf{h}(n) = [h_1(n), h_2(n), \cdots, h_L(n)]^T,$$

$v(n)$ is the measurement noise, and $L$ is the length of the impulse response. The estimated error is defined as

$$e(n) = d(n) - \mathbf{x}^T(n)\hat{\mathbf{h}}(n-1), \quad (2)$$

where $\hat{\mathbf{h}}(n)$ is the adaptive filter's coefficients.

The coefficient update of the family of PNLMS algorithms is [2]:

$$\hat{\mathbf{h}}(n) = \hat{\mathbf{h}}(n-1) + \frac{\mu \mathbf{G}(n-1)\mathbf{x}(n)e(n)}{\mathbf{x}^T(n)\mathbf{G}(n-1)\mathbf{x}(n) + \delta}, \quad (3)$$

where $\mu$ is the step-size, $\delta$ is the regularization parameter, and

$$\mathbf{G}(n-1) = diag[g_1(n-1), g_2(n-1), \cdots, g_L(n-1)]. \quad (4)$$

It should be noted that the step-size for the NLMS is the same for all filter coefficients: $\mathbf{G}(n-1) = \mathbf{I}_{L \times L}$, where $\mathbf{I}_{L \times L}$ is an $L \times L$ identity matrix. Meanwhile, the matrix for the family of PNLMS is defined as

$$g_l(n-1) = \frac{\gamma_l(n-1)}{\frac{1}{L}\sum_{i=1}^{L}\gamma_i(n-1)}, \quad (5)$$

where

$$\gamma_l = \max\left\{\rho \max\left\{q, F(|\hat{h}_1|), \cdots, F(|\hat{h}_L|)\right\}, F(|\hat{h}_l|)\right\}, \quad (6)$$

$F(|\hat{h}_l|)$ is specific to the algorithm, $q$ is a small positive value that prevents the filter coefficients $\hat{h}_l(n-1)$ from stalling when $\hat{\mathbf{h}}(0) = \mathbf{0}_{L \times 1}$ at initialization, and $\rho$, another small positive value, prevents the coefficients from stalling when they are much smaller than the largest coefficient [1]. The classical PNLMS employs step-sizes that are proportional to the magnitude of the estimated impulse response [2],

$$F(|\hat{h}_l(n-1)|) = |\hat{h}_l(n-1)|. \quad (7)$$

Instead of (5) and (6), the improved PNLMS (IPNLMS) algorithm proposed to use [10]

$$\gamma_l = (1-\alpha)\frac{\sum_{i=1}^{L}|\hat{h}_i|}{L} + (1+\alpha)|\hat{h}_l|, \quad (8)$$

and

$$g_l(n-1) = \frac{\gamma_l(n-1)}{\sum_{i=1}^{L}\gamma_i(n-1)} = \frac{(1-\alpha)}{2L} + \frac{(1+\alpha)|\hat{h}_l|}{2\sum_{i=1}^{L}|\hat{h}_i|}, \quad (9)$$

where $-1 \leq \alpha < 1$. IPNLMS behaves like NLMS when $\alpha = -1$ and PNLMS for $\alpha$ close to 1. In general, IPNLMS is a sum of two terms. The first term is an average of the absolute value of the coefficients taken from the estimated filter and the second is the absolute value of the coefficient itself. For most AEC/NEC applications, a good choice is $\alpha = 0, -0.5$, with which IPNLMS behaves better than either the NLMS or the PNLMS, regardless of the impulse response nature [10].

In next section, we will show that NLMS and PNLMS are all special cases of our proposed block-sparse PNLMS (BS-PNLMS). Meanwhile, we could further take advantage of the benefits of IPNLMS algorithms to improve the performance of the proposed BS-PNLMS algorithm.

### III. PROPOSED BS-PNLMS

The motivation behind the proposed family of the block-sparse proportionate algorithms is discussed at the beginning of this section, and then the proposed BS-PNLMS and BS-IPNLMS algorithms are presented next.

*A. Motivation of the Proposed BS-PNLMS*

A sparse impulse response is that in which a large percentage of the energy is distributed to only a few coefficients [1]. Several different types of sparse systems exist as indicated in Figure 1. The nonzero coefficients in a general sparse system (see Figure 1(a)) may be arbitrarily located. Meanwhile, there exists a special family known as either clustering-sparse systems or block-sparse systems [18]. For example, the network echo path is typically characterized by a bulk delay that is dependent on network loading, encoding, and jitter buffer delays. This results in an "active" region in the range of 8-12 *ms* duration, and the impulse response is dominated by "inactive" regions where coefficient magnitudes are close to zero [1]. The network echo response is a typical single-clustering sparse system (see Figure 1(b)). Satellite communication is an important modern application of echo cancellation. The impulse response of the echo path in satellite-linked communications consists of several long flat delay regions and disperse active regions. Such responses are representative of multi-clustering sparse systems. The waveform in a communication link that uses single-side band suppressed carrier modulation, contains both a relatively large near-end echo, characterized by a short time delay and a far-end echo that is smaller in amplitude but with a longer delay [20]. Therefore, the echo path impulse response is primarily characterized by two active regions that correspond to the near-end signal and the far-end signal echo (see Figure 1(c)). Considering the block-sparse characteristic of the sparse impulse responses, as in Figure 1(b) and Figure 1(c), the proportionate algorithm can be further improved by exploiting this special characteristic.

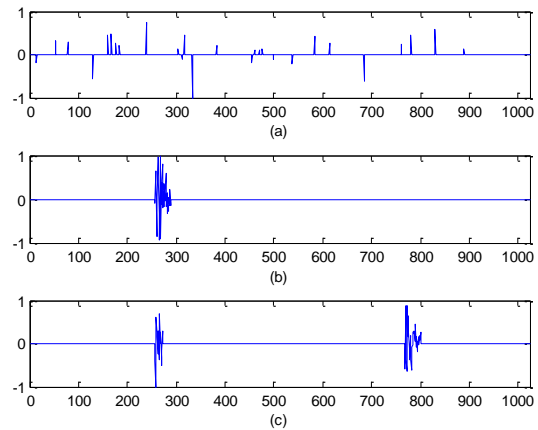

**Figure 1** Three types of sparse systems, (a) a general sparse system, (b) a one-cluster block-sparse system, and (c) a two-cluster block-sparse system.

It can be observed that an echo path, such as Figure 1(b), consists of the direct path and a few early reflections, which are almost always sparse, and the late reverberant part, which is always dispersive. The PB-IPNLMS algorithm splits the impulse response into two blocks and used two IPNLMS algorithms each with a different proportionate/non-proportionate factor for the two corresponding time-domain partitioned blocks [12].

However, the PB-IPNLMS in [12] depends on the assumption of one-cluster sparse system, which does not hold for the multi-clustering case as in Figure 1(c). Additional IPNLMS algorithms could be employed to extend the PB-IPNLMS to multi-cluster sparse system. However, this must depend on the priori information of the bulk delays in the multi-cluster sparse system, which is not necessarily the case in practice.

P. Loganathan et al. in [12] noted that distributing almost equal step-sizes for the dispersive block provides better steady-state performance, which agrees with the well-known fact that for the dispersive system, NLMS is preferred over PNLMS. Meanwhile, PNLMS is only beneficial when the impulse response is close to a delta function [9]. Therefore, the block-sparse proportionate NLMS (BS-PNLMS) algorithm is proposed to accelerate the convergence by combining the above two facts together. In BS-PNLMS, considering the fact that the block-sparse system is dispersive within each block, it is preferred to use NLMS within each block. Meanwhile, the idea of PNLMS can be applied to have the NLMS step-size for each block proportionate to its relative magnitude. More details are given in the following subsection.

*B. The Proposed BS-PNLMS Algorithm*

The proportionate NLMS algorithm can be deduced from a *basis pursuit* perspective [7]

$$\begin{aligned} \min &\quad \|\tilde{\mathbf{h}}(n)\|_1 \\ \text{subject to} &\quad d(n) = \mathbf{x}^T(n)\tilde{\mathbf{h}}(n), \end{aligned} \quad (10)$$

where $\tilde{\mathbf{h}}(n)$ is the correction component defined as [7]

$$\tilde{\mathbf{h}}(n) = \frac{\mathbf{G}(n-1)\mathbf{x}(n)d(n)}{\mathbf{x}^T(n)\mathbf{G}(n-1)\mathbf{x}(n)}.$$

Motivated by the observations in Section III.A, a family of proportionate adaptive algorithm for block-sparse system identification can be derived by replacing the $l_1$ norm optimization target in (10) with the following $l_{2,1}$ norm defined as

$$\|\tilde{\mathbf{h}}\|_{2,1} = \left\| \begin{bmatrix} \|\tilde{\mathbf{h}}_{[1]}\|_2 \\ \|\tilde{\mathbf{h}}_{[2]}\|_2 \\ \vdots \\ \|\tilde{\mathbf{h}}_{[N]}\|_2 \end{bmatrix} \right\|_1 = \sum_{i=1}^{N} \|\tilde{\mathbf{h}}_{[i]}\|_2, \quad (11)$$

where $\|\tilde{\mathbf{h}}_{[i]}\|_2 = \sqrt{\tilde{\mathbf{h}}_{[i]}^T \tilde{\mathbf{h}}_{[i]}}$, $\tilde{\mathbf{h}}_{[i]} = [\tilde{h}_{(i-1)P+1} \tilde{h}_{(i-1)P+1} \cdots \tilde{h}_{iP}]^T$, and $P$ is a predefined group partition size parameter. The following convex target could be minimized with a constraint on the linear system of equations:

$$\begin{aligned} \min &\quad \|\tilde{\mathbf{h}}(n)\|_{2,1} \\ \text{subject to} &\quad d(n) = \mathbf{x}^T(n)\tilde{\mathbf{h}}(n). \end{aligned} \quad (12)$$

The Lagrange multiplier can be used to derive the proposed block-sparse proportionate NLMS algorithm [6]-[7]. The derivative of the $l_{2,1}$ norm in (11), with respect to the weight vector, is

$$\frac{\partial \|\tilde{\mathbf{h}}(n)\|_{2,1}}{\partial \tilde{\mathbf{h}}(n)} = \left[ \frac{\partial \sum_{i=1}^{N} \|\tilde{\mathbf{h}}_{[i]}\|_2}{\partial \tilde{h}_1}, \frac{\partial \sum_{i=1}^{N} \|\tilde{\mathbf{h}}_{[i]}\|_2}{\partial \tilde{h}_2}, \cdots, \frac{\partial \sum_{i=1}^{N} \|\tilde{\mathbf{h}}_{[i]}\|_2}{\partial \tilde{h}_L} \right], \quad (13)$$

in which

$$\frac{\partial \sum_{i=1}^{N} \|\tilde{\mathbf{h}}_{[i]}\|_2}{\partial \tilde{h}_k} = \frac{\partial \|\tilde{\mathbf{h}}_{[j]}\|_2}{\partial \tilde{h}_k} = \frac{\tilde{h}_k}{\|\tilde{\mathbf{h}}_{[j]}\|_2}, \quad (14)$$
$$(j-1)P+1 \le k \le jP.$$

The update equation for the proposed BS-PNLMS is

$$\hat{\mathbf{h}}(n) = \hat{\mathbf{h}}(n-1) + \frac{\mu \mathbf{G}(n-1)\mathbf{x}(n)e(n)}{\mathbf{x}^T(n)\mathbf{G}(n-1)\mathbf{x}(n) + \delta}, \quad (15)$$

where

$$\mathbf{G}(n-1) = \\ diag\left[ \|\hat{\mathbf{h}}_{[1]}\|_2 \mathbf{1}_P, \|\hat{\mathbf{h}}_{[2]}\|_2 \mathbf{1}_P, \cdots, \|\hat{\mathbf{h}}_{[N]}\|_2 \mathbf{1}_P \right], \quad (16)$$

and $\mathbf{1}_P$ is a $P$-length row vector of all ones. Equation (15) is the same as the traditional PNLMS, except that here the block-sparse definition of $\mathbf{G}(n-1)$ is used in (16). In a manner similar to (4)-(6) in PNLMS to prevent stalling issues, the proposed BS-PNLMS does so as

$$\mathbf{G}(n-1) = \\ diag\left[ g_1(n-1)\mathbf{1}_P, g_2(n-1)\mathbf{1}_P, \cdots, g_N(n-1)\mathbf{1}_P \right], \quad (17)$$

where

$$g_i(n-1) = \frac{\gamma_i}{\frac{1}{N}\sum_{l=1}^{N}\gamma_l}, \quad (18)$$

and

$$\gamma_i = \max\left\{ \rho \max\left\{ q, \|\hat{\mathbf{h}}_{[1]}\|_2, \cdots, \|\hat{\mathbf{h}}_{[N]}\|_2 \right\}, \|\hat{\mathbf{h}}_{[i]}\|_2 \right\}. \quad (19)$$

The traditional PNLMS and NLMS algorithms can each be easily verified as special cases of the proposed BS-PNLMS. If $P$ is equal to 1, the mixed $l_{2,1}$ norm in (11) is equivalent to the $l_1$ norm in (10), which is the classical *basis pursuit* based PNLMS algorithm [7]. Meanwhile, if $P$ is chosen as $L$, the mixed $l_{2,1}$ norm in (13) is the same as the $l_2$ norm and BS-PNLMS then becomes the traditional NLMS [7]. Therefore, the BS-PNLMS is a generalization of NLMS and PNLMS.

### C. Extension to the BS-IPNLMS Algorithm

Meanwhile, in order to further improve the robustness of the proposed BS-PNLMS algorithm to both sparse and dispersive impulse responses, an improved BS-PNLMS (BS-IPNLMS) algorithm is proposed using the similar idea of IPNLMS algorithm

$$\gamma_l = (1-\alpha)\frac{\sum_{i=1}^{N}\|\hat{\mathbf{h}}_{[i]}\|_2}{N} + (1+\alpha)\|\hat{\mathbf{h}}_{[l]}\|_2 \quad (20)$$

$$g_l(n) = \frac{\gamma_l(n)}{P\sum_{i=1}^{N}\gamma_i(n)} = \frac{(1-\alpha)}{2L} + \frac{(1+\alpha)\|\hat{\mathbf{h}}_{[l]}\|_2}{2P\sum_{i=1}^{N}\|\hat{\mathbf{h}}_{[i]}\|_2}. \quad (21)$$

This section is concluded with a brief discussion about the proposed BS-PNLMS and BS-IPNLMS algorithms. Unlike the PB-IPNLMS, the proposed BS-PNLMS and BS-IPNLMS algorithms only require prior information about the *length* of the active regions to determine the group size, which are usually known for both the NEC and the satellite link channels, etc., and not their actual *locations*. The BS-PNLMS could be interpreted as transferring the block-sparse system into a multi-delta system in the coefficient space to fully benefit from PNLMS. However, if the impulse system is dispersive, or the group size is much smaller than the actual block size in the impulse response, the BS-IPNLMS could outperform both the PNLMS and the BS-PNLMS, as well. The details of the proposed BS-PNLMS and BS-IPNLMS algorithms are summarized in Table 1. The superior performance of BS-PNLMS, and BS-IPNLMS over NLMS, PNLMS, and IPNLMS will be demonstrated in the simulations of Section V.

## IV. COMPUTATIONAL COMPLEXITY

The computational complexity of BS-PNLMS and BS-IPNLMS algorithms is compared with traditional NLMS, PNLMS and IPNLMS algorithms in Table 2 in terms of the total number of additions (A), multiplications (M), divisions (D), comparisons (C), square roots (Sqrt) and memory words (MW), needed per sample. The additional computational complexity for the BS-PNLMS family arises from the computation of the $l_2$ norm of the block responses using the square root operations. The complexity of the square root can be reduced through the use of a look up table or a Taylor series expansion [22]. Meanwhile, it should be noted that the "comparison operations" and the required memory words for the family of BS-PNLMS are decreased from that of PNLMS. Finally, the computational complexity of the proposed block-sparse family algorithms is also related to the number of groups, $N$, where $N = L/P$.

## V. SIMULATION RESULTS

Simulations were conducted to evaluate the performance of the proposed BS-PNLMS and BS-IPNLMS algorithms. The algorithms were tested using zero mean white Gaussian noise (WGN), colored noise and speech signals at sampling rate 8 KHz. The WGN was filtered through a first order system with a pole at 0.8 to generate the colored input signals. An independent WGN was added to the system's background at a signal-to-noise ratio (SNR) of 30dB. The regularization parameter for NLMS was $\delta_{NLMS} = 0.01$, and the regularization parameters for PNLMS, BS-PNLMS, IPNLMS, and BS-IPNLMS were $\delta_{NLMS}/L$ according to [19]. The values of $\alpha$ used for both the IPNLMS and the BS-IPNLMS algorithms were 0. For both the PNLMS and the BS-PNLMS algorithms, $\rho = 0.01$, and $q = 0.01$.

The convergence state of adaptive filter was evaluated with the normalized misalignment defined as
$10\log_{10}(\|\mathbf{h}-\hat{\mathbf{h}}\|_2^2 / \|\mathbf{h}\|_2^2)$.

**Table 1.** The Block-Sparse Algorithms

Initializations:
$$\hat{\mathbf{h}}(n) = \mathbf{0}_{L\times 1}, \ N = L/P$$

General Computations:
$$e(n) = d(n) - \mathbf{x}^T(n)\hat{\mathbf{h}}(n-1)$$
$$\mathbf{G}(n-1) = diag[g_1\mathbf{1}_P, g_2\mathbf{1}_P, \cdots, g_N\mathbf{1}_P]$$
$$\hat{\mathbf{h}}(n) = \hat{\mathbf{h}}(n-1) + \frac{\mu \mathbf{G}(n-1)\mathbf{x}(n)e(n)}{\mathbf{x}^T(n)\mathbf{G}(n-1)\mathbf{x}(n) + \delta}$$
for $i = 1, 2, \cdots, N$
$$\|\hat{\mathbf{h}}_{[i]}\|_2 = \sqrt{\sum_{k=1}^{P} \hat{h}_{(i-1)P+k}^2}$$
end for

BS-PNLMS:
for $i = 1, 2, \cdots, N$
$$\gamma_i = \max\left\{\rho \max\left\{q, \|\hat{\mathbf{h}}_{[1]}\|_2, \cdots, \|\hat{\mathbf{h}}_{[N]}\|_2\right\}, \|\hat{\mathbf{h}}_{[i]}\|_2\right\}$$
$$g_i(n) = \frac{\gamma_i}{\frac{1}{N}\sum_{l=1}^{N}\gamma_l}$$
end for

BS-IPNLMS:
for $i = 1, 2, \cdots, N$
$$g_l(n) = \frac{(1-\alpha)}{2L} + \frac{(1+\alpha)\|\hat{\mathbf{h}}_{[l]}\|_2}{2P\sum_{i=1}^{N}\|\hat{\mathbf{h}}_{[i]}\|_2}$$
end for

**Table 2.** Computational complexity of the algorithms' coefficient updates – Addition (A), Multiplication (M), Division (D), Comparison (C), Square Root (Sqrt) and Memory Word (MW).

| Algorithm | A | M | D | C | Sqrt | MW |
|---|---|---|---|---|---|---|
| NLMS | 2$L$+3 | 2$L$+3 | 1 | 0 | 0 | 4$L$+7 |
| PNLMS | 4$L$+2 | 5$L$+4 | 2 | 2$L$ | 0 | 8$L$+11 |
| BS-PNLMS | 4$L$-1 | 6$L$+3 | 2 | $N$+1 | $N$ | 5$L$+3$N$+11 |
| IPNLMS | 5$L$+2 | 6$L$+2 | 4 | $L$-1 | 0 | 8$L$+11 |
| BS-IPNLMS | 4$L$+$N$-1 | 6$L$+$N$+1 | 2 | 0 | $N$ | 5$L$+3$N$+11 |

In all the simulations except for the ones in section V.C, the length of the unknown system throughout the simulation was $L = 1024$, and the adaptive filter had the same length. A 32 taps impulse response in Figure 1 (b) with a single cluster of nonzero coefficients at [257, 288] was used. In order to compare the tracking ability for different algorithms, an echo path change was incurred at 40000 sample by switching to the two-clusters response located at [257, 272] (16 taps) and [769, 800] (32 taps) as illustrated in in Figure 1 (c). All the algorithms were simulated for five times and averaged in order to evaluate their performance.

### A. Effect of P on the Performance of BS-PNLMS

In order to demonstrate the effect of $P$, the performance of the proposed BS-PNLMS was tested for different group sizes $P$ (4, 16, 32, and 64) separately. Meanwhile, the performance of NLMS, which is the same as BS-PNLMS with $P=1024$, and PNLMS (the same as BS-PNLMS with $P=1$) algorithms were also included. In the first simulation in Figure 2 (a), the input was WGN, and the step-size $\mu$ was set to 0.1. The simulation results for a colored input signal and speech input signal are illustrated in Figure 2 (b) and Figure 2 (c) separately, where the step-sizes were $\mu=0.2$ for both the colored input and the speech input. Meanwhile, the remaining parameters for the three simulations were the same.

Simulation results in Figure 2 indicate that the group size $P$ should be chosen properly in order to gain better performance than either the NLMS or the PNLMS. Due to the fact that there are a total 32 taps in the single-cluster impulse response, it is reasonable that the group size larger than 32 will likely degrade the performance before the echo path change. Meanwhile, there are two clusters with length 16 taps separately in the two-cluster impulse response, and the group size should be smaller than 16. Because the groups are evenly spaced, the actual block could have been split into multiple groups too. Therefore, the group size should be smaller than the length of cluster's actual minimum size in the impulse response. The cluster's size is typically known in real-world applications. For example, the NEC's "active" region is in the range of 8-12 $ms$ duration [1]. If the group size is significantly larger than the cluster size of block-sparse system, the convergence speed will become worse than the traditional PNLMS. This fact is intuitive, considering that NLMS, which uses $P=1024$, converges slower than PNLMS with $P=1$ for a block-sparse system. Thus, both NLMS and PNLMS represent extreme cases. The NLMS algorithm should be chosen when the unknown system is dispersive, i.e. the cluster size is the length of the full filter, and when the unknown system is generally sparse as illustrated in Figure 1(a), PNLMS should be used because the cluster size is 1.

### B. Convergence Performance of BS-PNLMS and BS-IPNLMS for Block-Sparse Systems

The performances of NLMS, PNLMS, IPNLMS, proposed BS-PNLMS with $P=16$ and the proposed BS-IPNLMS with $P=4$ were compared for the two block-sparse systems in Figure 3.

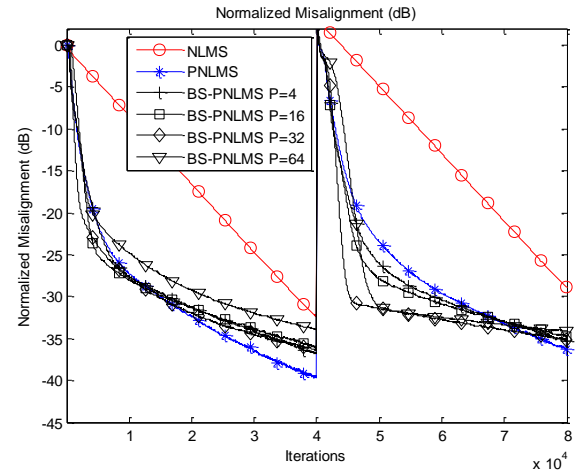

(a) WGN input with $\mu=0.1$

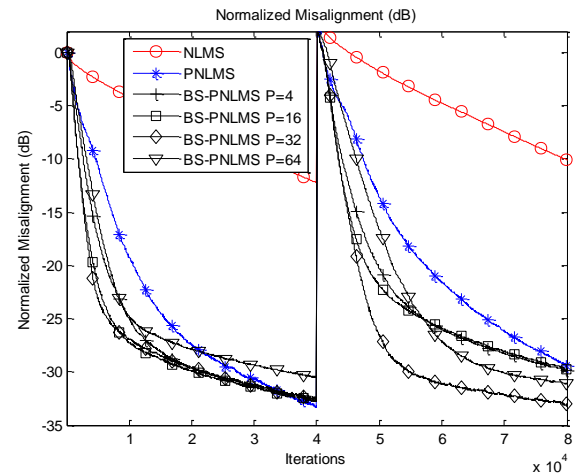

(b) Colored noise input with $\mu=0.2$

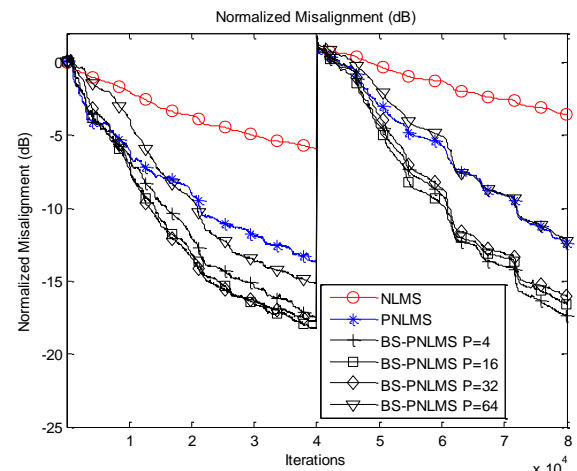

(c) Speech input with $\mu=0.2$

**Figure 2** Comparison of the BS-PNLMS algorithms with different group sizes for block-sparse systems in Figure 1 (b) and Figure 1 (c) at SNR=30dB: (a) white, (b) colored noise and (c) speech input signals.

The WGN was used as the input signal in Figure 3 (a) with the step-sizes as $\mu_{NLMS} = \mu_{PNLMS} = 0.1$, and $\mu_{BS-PNLMS} = \mu_{BS-IPNLMS} = 0.1$. The simulation results for the colored and speech input are illustrated in Figure 3 (b) and Figure 3 (c), where $\mu_{NLMS} = \mu_{PNLMS} = 0.2$, and $\mu_{BS-PNLMS} = \mu_{BS-IPNLMS} = 0.2$.

The proposed BS-PNLMS algorithm provides faster convergence rate and tracking ability than either the NLMS or the traditional PNLMS algorithms for the block-sparse impulse responses. Meanwhile, the convergence rate of BS-IPNLMS outperformed both the NLMS and the IPNLMS algorithms.

It is interesting to observe that the BS-PNLMS algorithm outperformed the BS-IPNLMS algorithm. This is due to fact that the two block-sparse systems in Figure 1 (b) and Figure 1 (c) are very sparse. Meanwhile, the BS-PNLMS transformed them into highly sparse systems with only 2 or 3 non-zero elements which fully benefits from PNLMS. Meanwhile, the benefits of BS-IPNLMS for the dispersive impulse responses will be demonstrated in the next subsection.

### C. Convergence Performance of BS-PNLMS and BS-IPNLMS for the Acoustic Echo Path and a Random Dispersive System

In order to verify the performance of the proposed BS-IPNLMS algorithm for dispersive impulse response, simulations were conducted to compare the performances of NLMS, PNLMS, IPNLMS, the proposed BS-PNLMS with $P=16$, and the proposed BS-IPNLMS with $P=16$. An echo path change was incurred at 40000 samples by switching from a 512 taps measured acoustic echo path in Figure 4 (a) to a random impulse response in Figure 4 (b). The simulation results for WGN, colored noise and speech input signals are illustrated in Figure 5.

The step-size parameters were $\mu_{NLMS} = \mu_{PNLMS} = 0.2$, $\mu_{BS-PNLMS} = \mu_{BS-IPNLMS} = 0.2$ for the WGN input, and $\mu_{NLMS} = \mu_{PNLMS} = 0.4$, $\mu_{BS-PNLMS} = \mu_{BS-IPNLMS} = 0.4$ for both the colored noise and the speech input signals.

It can be observed that the BS-IPNLMS algorithm outperformed the BS-PNLMS algorithm for both the acoustic echo path and the random dispersive impulse response. Meanwhile, both BS-PNLMS and BS-IPNLMS work better than the traditional PNLMS algorithm for the random dispersive impulse responses.

It should be noted that, neither the acoustic echo path nor the random dispersive impulse response are typical block-sparse impulse systems, therefore, the family of BS-IPNLMS should be used to obtain better performance instead of the BS-PNLMS algorithms.

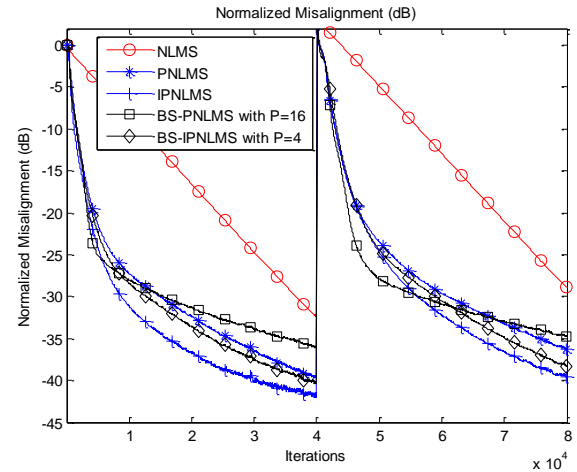

(a) $\mu_{NLMS} = \mu_{PNLMS} = 0.1$, $\mu_{BS-PNLMS} = \mu_{BS-IPNLMS} = 0.1$

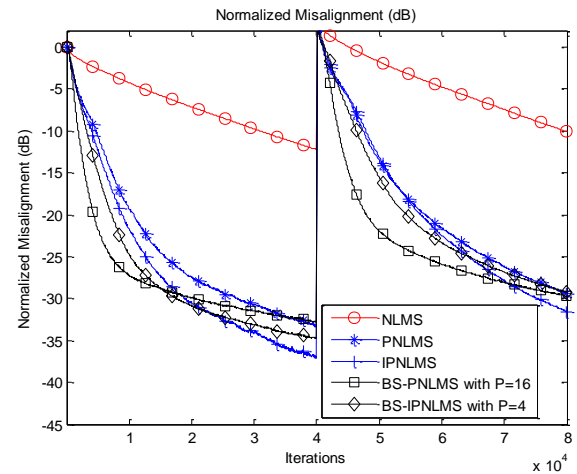

(b) $\mu_{NLMS} = \mu_{PNLMS} = 0.2$, $\mu_{BS-PNLMS} = \mu_{BS-IPNLMS} = 0.2$

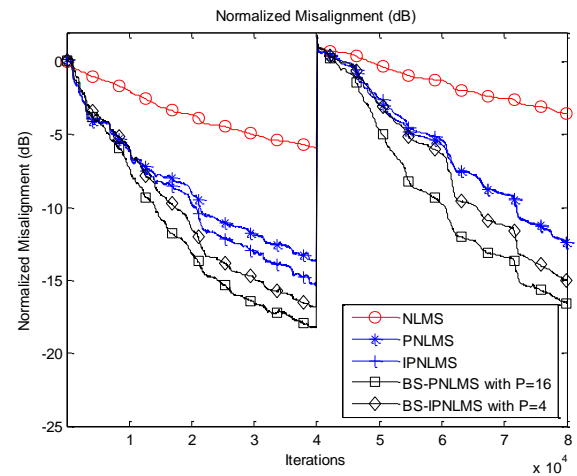

(c) $\mu_{NLMS} = \mu_{PNLMS} = 0.2$, $\mu_{BS-PNLMS} = \mu_{BS-IPNLMS} = 0.2$

**Figure 3** Comparison of NLMS, PNLMS, IPNLMS, BS-PNLMS and BS-IPNLMS algorithms for block-sparse systems in Figure 1 (b) and Figure 1 (c) at SNR=30dB: (a) WGN input, (b) colored noise and (c) speech input signals.

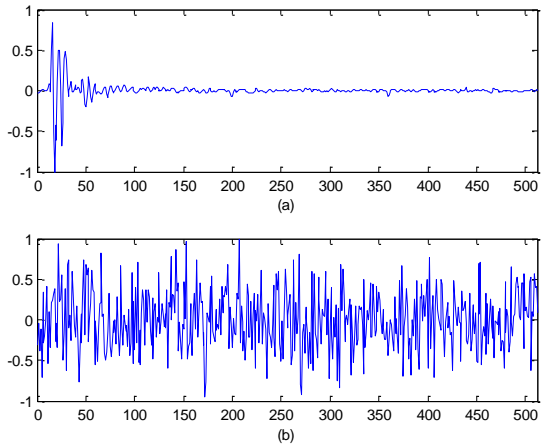

**Figure 4** Two impulse responses (a) a measured quasi-sparse acoustic echo path, (b) a random dispersive impulse response.

## VI. CONCLUSION AND FUTURE WORK

A new family of proportionate algorithms for block-sparse system identification (known as BS-PNLMS and BS-IPNLMS) were proposed. These algorithms were based on the optimization of a mixed $l_{2,1}$ norm of the adaptive filter's coefficients. The computational complexities of the proposed algorithms were presented. Simulation results demonstrated that, the new BS-PNLMS algorithm outperforms the NLMS, PNLMS and IPNLMS algorithms for the block-sparse system, and the new BS-IPNLMS algorithm is more preferred for the dispersive system.

This block-sparse proportionate idea proposed in this paper could be further extended to many other proportionate algorithms, including proportionate affine projection algorithm (PAPA) [23], proportionate affine projection sign algorithm (PAPSA) [24], and their corresponding low complexity implementations [25]-[26] etc. The proof of convergence for the proposed BS-PNLMS and BS-IPNLMS algorithms can also be part of the future work. Finally, it will be interesting to explore the variable and non-uniform group split to further improve the performance of the BS-PNLMS and the BS-IPNLMS algorithms.

## ACKNOWLEDGEMENT

The authors would like to thank the Associate Editor and the reviewers for the valuable comments and suggestions.

## REFERENCES

[1] K. Wagner and M. Doroslovački, *Proportionate-type Normalized Least Mean Square Algorithms*: John Wiley & Sons, 2013.

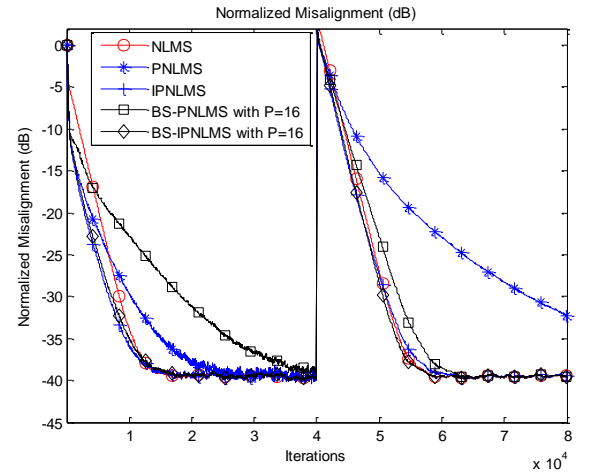

(a) $\mu_{NLMS} = \mu_{PNLMS} = 0.2$, $\mu_{BS-PNLMS} = \mu_{BS-IPNLMS} = 0.2$

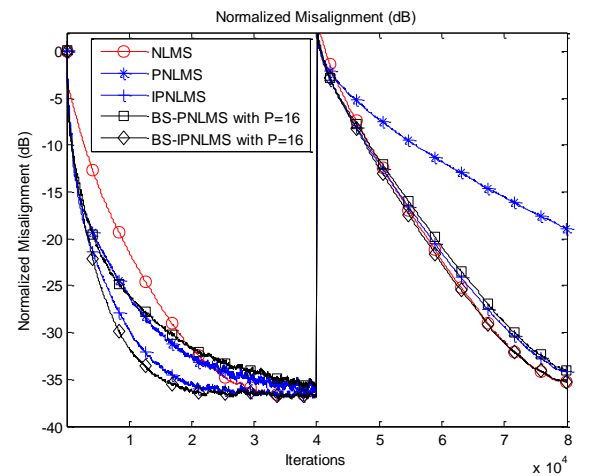

(b) $\mu_{NLMS} = \mu_{PNLMS} = 0.4$, $\mu_{BS-PNLMS} = \mu_{BS-IPNLMS} = 0.4$

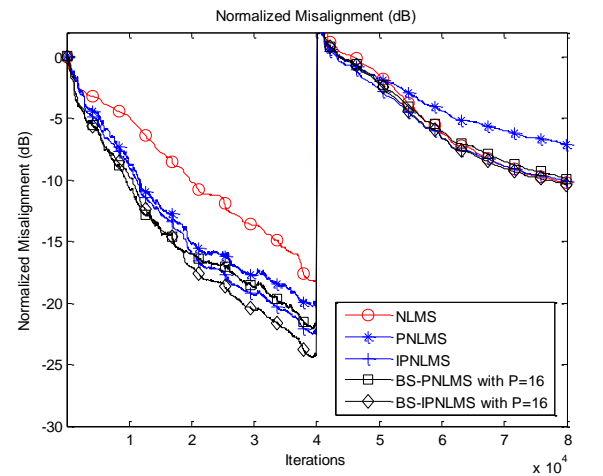

(c) $\mu_{NLMS} = \mu_{PNLMS} = 0.4$, $\mu_{BS-PNLMS} = \mu_{BS-IPNLMS} = 0.4$

**Figure 5** Comparison of NLMS, PNLMS, IPNLMS, BS-PNLMS and BS-IPNLMS algorithms for acoustic echo path and dispersive system in Figure 4 and SNR=30dB: (a) WGN input (b) colored noise and (c) speech input.

[2] D. L. Duttweiler, "Proportionate normalized least-mean-squares adaptation in echo cancelers," *Speech and Audio Processing, IEEE Transactions on,* vol. 8, no. 5, pp. 508-518, 2000.

[3] H. Deng and M. Doroslovački, "Improving convergence of the PNLMS algorithm for sparse impulse response identification," *Signal Processing Letters, IEEE,* vol. 12, no. 3, pp. 181-184, 2005.

[4] H. Deng and M. Doroslovački, "Proportionate adaptive algorithms for network echo cancellation," *Signal Processing, IEEE Transactions on,* vol. 54, no. 5, pp. 1794-1803, 2006.

[5] K. Wagner, M. Doroslovački, and H. Deng, "Convergence of proportionate-type LMS adaptive filters and choice of gain matrix," *40th Asilomar Conference on Signals, Systems and Computers*, 2006 (ACSSC '06), Pacific Grove, CA, pp. 243–247, November 2006.

[6] C. Paleologu, J. Benesty and S. Ciochină, "An improved proportionate NLMS algorithm based on $l_0$ norm," *IEEE ICASSP*, pp. 309-312, 2010.

[7] J. Benesty, C. Paleologu, and S. Ciochină, "Proportionate adaptive filters from a basis pursuit perspective," *Signal Processing Letters, IEEE,* vol. 17, no. 12, pp. 985-988, 2010.

[8] J. Liu and S. L. Grant, "A generalized proportionate adaptive algorithm based on convex optimization," in *Proc. Signals and Information Processing (ChinaSIP), 2014 IEEE China Summit & International Conference on*, pp. 748-752, 2014.

[9] S. L. Gay, "An efficient, fast converging adaptive filter for network echo cancellation," *Conference Record of the 32nd Asilomar Conference on Signals, Systems and Computers*, Pacific Grove, CA, vol. 1, pp. 394–398, November 1998.

[10] J. Benesty and S. L. Gay, "An improved PNLMS algorithm," *IEEE ICASSP*, pp. 1881-1884, 2002.

[11] J. Cui, P. Naylor, and D. Brown, "An improved IPNLMS algorithm for echo cancellation in packet-switched networks," *IEEE International Conference on Acoustics, Speech, and Signal Processing*, 2004 (ICASSP '04), vol. 4, Montreal, Quebec, Canada, pp. 141–144, May 2004.

[12] P. Loganathan, E. A. P. Habets and P. A. Naylor, "A partitioned block proportionate adaptive algorithm for acoustic echo cancellation," *Proc. of the APSIPA Annual Summit and Conference* 2010, Biopolis, Singapore, Dec 2010.

[13] F. D. C. De Souza, O. J. Tobias, R. Seara *et al.*, "A PNLMS algorithm with individual activation factors," *Signal Processing, IEEE Transactions on,* vol. 58, no. 4, pp. 2036-2047, 2010.

[14] Y. Chen, Y. Gu and A. O. Hero III, "Sparse LMS for system identification," *IEEE ICASSP*, pp. 3125-3128, 2009.

[15] Y. Gu, J. Jin, and S. Mei, "l0 norm constraint LMS algorithm for sparse system identification," *Signal Processing Letters, IEEE,* vol. 16, no. 9, pp. 774-777, 2009.

[16] J. Liu and S. L. Grant, "A new variable step-size zero-point attracting projection algorithm," in *Proc. Signals, Systems and Computers, 2013 Asilomar Conference*, pp. 1524-1528, 2013.

[17] J. Liu and S. L. Grant, "An improved variable step-size zero-point attracting projection algorithm," *Acoustics, Speech and Signal Processing (ICASSP), 2015 IEEE International Conference on* , pp.604-608, 19-24 April 2015

[18] S. Jiang and Y. Gu, "Block-Sparsity-Induced Adaptive Filter for Multi-Clustering System Identification," in *Signal Processing, IEEE Transactions on* , vol.63, no.20, pp.5318-5330, Oct.15, 2015

[19] J. Benesty, C. Paleologu, and S. Ciochină, "On regularization in adaptive filtering," *Audio, Speech, and Language Processing, IEEE Transactions on,* vol. 19, no. 6, pp. 1734-1742, 2011.

[20] P. A. Marques, F. M. Sousa, and J. Leitao, "A DSP based long distance echo canceller using short length centered adaptive filters," In *Acoustics, Speech, and Signal Processing, 1997. ICASSP-97., 1997 IEEE International Conference on*, vol. 3, pp. 1885-1888. IEEE, 1997.

[21] P. Loganathan, A. W. Khong, and P. Naylor, "A class of sparseness-controlled algorithms for echo cancellation," *Audio, Speech, and Language Processing, IEEE Transactions on,* vol. 17, no. 8, pp. 1591-1601, 2009.

[22] D. Liu, *Embedded DSP processor design: application specific instruction set processors*: Morgan Kaufmann, 2008.

[23] T. Gansler, J. Benesty, S. L. Gay, and M. Sondhi. "A robust proportionate affine projection algorithm for network echo cancellation," In *Acoustics, Speech, and Signal Processing, 2000. ICASSP'00. Proceedings. 2000 IEEE International Conference on*, vol. 2, pp. II793-II796. IEEE, 2000.

[24] Z. Yang, Y. R. Zheng, and S. L. Grant, "Proportionate affine projection sign algorithms for network echo cancellation," *Audio, Speech, and Language Processing, IEEE Transactions on,* vol. 19, no. 8, pp. 2273-2284, 2011.

[25] C. Paleologu, S. Ciochină, and J. Benesty, "An efficient proportionate affine projection algorithm for echo cancellation," *Signal Processing Letters, IEEE* 17, no. 2 (2010): 165-168.

[26] F. Albu and H. K. Kwan, "Memory improved proportionate affine projection sign algorithm," *Electronics letters* 48, no. 20 (2012): 1279-1281.

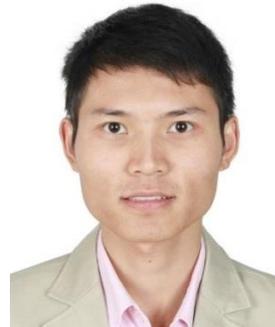

**Jianming Liu** (S'14) received his B.S. degree in Electronic Science and Technology from Shandong University, Jinan, China, in 2005, and M.S. in Electrical and Communication Engineering from Tsinghua University, Beijing, China, in 2009. From 2010 to 2011, he worked as an Audio Design Engineer in Nokia, Beijing, China and he is currently pursuing the Ph.D. degree in Electrical and Computer Engineering at Missouri University of Science and Technology, Rolla, USA. His research interests include adaptive filtering, echo cancellation and audio signal processing.

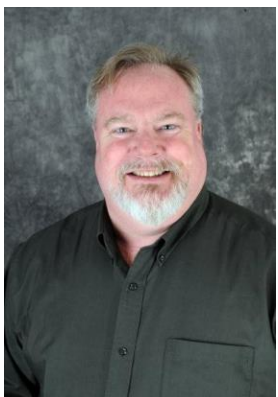 **Steven L. Grant** is also known as Steven L. Gay. He received his B.S in electrical engineering from the Missouri University of Science and Technology in 1979. In 1980, he joined Bell Labs in Whippany, New Jersey and received his M.S. in electrical engineering from Caltech. He worked in Bell Labs development organizations until 1991 when he joined the Bell Labs Acoustics and Speech Research Department in Murray Hill. His work there focused on acoustic signal processing. While there, he completed his Ph.D. in electrical engineering at Rutgers University in 1994. In 2001, he became the technical manager of the acoustics research group. From 2002 to 2004, he worked at the MIT Lincoln Lab on nonlinear equalization and sonar research. He is now the Wilkens Missouri Telecommunications Associate Professor of Electrical Engineering at Missouri S&T. He was co-chair and general chair of the 1999 and 2008 International Workshop on Acoustic Echo and Noise Control (IWAENC) and serves on its standing committee. He has edited two books and co-authored another on DSP algorithms and applications. He has also served as associate editor of IEEE Signal Processing Letters and IEEE Transactions on Signal Processing.